# Towards semantic and affective coupling in emotionally annotated databases

M. Horvat[1], S. Popović[1] and K. Ćosić[1]
[1] Faculty of Electrical Engineering and Computing, University of Zagreb
Department of Electric Machines, Drives and Automation
Zagreb, Croatia
marko.horvat2@fer.hr

**Abstract** - Emotionally annotated databases are repositories of multimedia documents with annotated affective content that elicit emotional responses in exposed human subjects. They are primarily used in research of human emotions, attention and development of stress-related mental disorders. This can be successfully exploited in larger processes like selection, evaluation and training of personnel for occupations involving high stress levels. Emotionally annotated databases are also used in multimodal affective user interfaces to facilitate richer and more intuitive human-computer interaction. Multimedia documents in emotionally annotated databases must have maximum personal ego relevance to be the most effective in all these applications. For this reason flexible construction of subject-specific of emotionally annotated databases is imperative. But current construction process is lengthy and labor intensive because it inherently includes an elaborate tagging experiment involving a team of human experts. This is unacceptable since the creation of new databases or modification of the existing ones becomes slow and difficult. We identify a positive correlation between the affect and semantics in the existing emotionally annotated databases and propose to exploit this feature with an interactive relevance feedback for a more efficient construction of emotionally annotated databases. Automatic estimation of affective annotations from existing semantics enhanced with information refinement processes may lead to an efficient construction of high-quality emotionally annotated databases.

## I. INTRODUCTION

Coupling of semantics and affect in multimedia documents is an interaction between description of object and events represented in a document and affect which they are eliciting. In other words, coupling refers to how the content of a multimedia document, and its perceived meaning, is related to emotions, sentiments and moods a human subject is experiencing when exposed to the document (i.e. multimedia stimulus). The underlying assumption is that similar semantics will provoke similar emotions in standard subjects [1][2]. The function describing functional dependency between semantics and emotion is surjective because two or more distinct semantics may trigger the same emotion.

We postulate the semantics-affect coupling rule in emotionally annotated databases and apply the coupling to a construction process of these databases. The coupling must be used with a premise of statistically uniform population of subjects which server as an annotation baseline. Although limiting in terms of experimentation this is necessary and also sufficient to define the coupling rule in its minimal form. In our work we rely on International Affective Picture System (IAPS) [3]. This multimedia database with almost 1000 different images is constructed on a statistically significant number of affective responses from a controlled population of subjects. The database has so far been used in a wide range of research (examples [4][5][6]) and as such provides an adequate source of quality data for identification of semantics-emotion pairs and their mutual coupling.

Determinate dependency between semantics and emotions can be pragmatically exploited in construction of emotionally annotated databases. This is very important since all contemporary emotionally annotated databases have been constructed manually. The coupling enables faster and less job-intensive identification of emotion present in multimedia based on known semantics, which in turn allows algorithms for semi-automatic annotations of stimuli and supported construction of emotionally annotated databases.

The layout of the papers is as follows: the next chapter describes emotionally annotated databases, their format and content. The third chapter deals with the functional definition of the semantic-affective coupling. The fourth chapter gives two examples of the coupling in IAPS. The fifth chapters explains how emotionally annotated databases can be constructed using information refinement with the coupling. Since all contemporary emotionally annotated databases have been constructed manually by different groups of experts, this chapter is important in clarifying why the coupling is a useful feature in construction of emotionally annotated databases and how it can be exploited to make the construction process simpler, faster and less work intensive. Finally, the conclusion is given in the final chapter.

## II. EMOTIONALLY ANNOTATED DATABASES

Emotionally annotated database or more precisely affective annotated multimedia database is a repository of semantically and affectively indexed multimedia documents such as [3][7][8][9][10][11][12]. Apart from

This research has been partially supported by the Croatian Ministry of Science, Education and Sports.

still images, sounds, video-clips some emotionally annotated databases may contain text in a form of individual words or short sentences [9][10]. Emotionally annotated databases contain information about affect and semantics of every stored document. This specific trait distinguishes them from other multimedia repositories. Specifically, emotionally annotated databases contain information about affective values of multimedia documents, i.e. the expected emotion or affect which will be induced in a human subject when exposed to the document. Because affectively annotated multimedia induces or stimulates emotional response the multimedia documents are also called multimedia stimuli or just stimuli.

Multimedia documents may provoke not just one but several emotional responses with different intensity [2]. In computer systems structure of emotion is categorized in two models: discrete and dimensional [13]. Although there is still some controversy on which approach is the best, in terms of computer systems implementation both models are simple and easy-to-use. Additionally, several other ad-hoc approaches in emotion state categorization exist such as those for expression of sentiment [11] and description of emotions' taxonomy [14].

Affective annotation of multimedia is a lengthy and laborious task that always includes a psychological experiment conducted with a statistically relevant group of participants. These experiments vary from the simple, such as the validation of a universal facial expression in a picture [8], to the elaborated which may include hundreds of participants [3]. To establish a baseline response all answers in the experiment are aggregated and statistically processed. Since they follow Gaussian distribution the affective annotations are described with the mean value and standard deviation. The mean represents the baseline and standard deviation the level of confidence in the aggregated baseline.

In the current form all emotionally annotated databases have very simple structures usually consisting of a file repository and a manifesto file that describes the content of the belonging repository [15]. The manifesto is comma-separated values (CSV) formatted textual file with attributes such as unique identifier, semantic descriptor and eliciting affect for each multimedia document in the repository. In the case of IAPS the unique identifier is the name of the stimulus file, e.g. 5200.jpg, 5201.jpg, etc. Through the combination of unique folder paths and unique file names within them the stimuli obtain their URIs (Uniform Resource Identificator) which allows their retrieval.

Some emotionally annotated databases like Geneva Affective PicturE Database (GAPED) [12], NimStim Face Stimulus Set [8] or Karolinska Directed Emotional Faces (KDEF) [16] use named files to describe stimuli valence and semantic content. These databases do not have a manifesto file but rather their affect and semantics are simple enough to be implicitly described with unique names of pictures or folders they are stored in. This is the simplest database structure and only minimally sufficient in conveying meaning of stimuli. In these cases it is expected that database user must look over all stimuli and selected them manually before using them. If the database does not contain many stimuli per single level of semantic taxonomy manual document retrieval from these databases is still manageable. For example, GAPED – currently the latest emotionally annotated database developed – has 730 negative, neutral and positive pictures but only up to 159 stimuli in a single named folder.

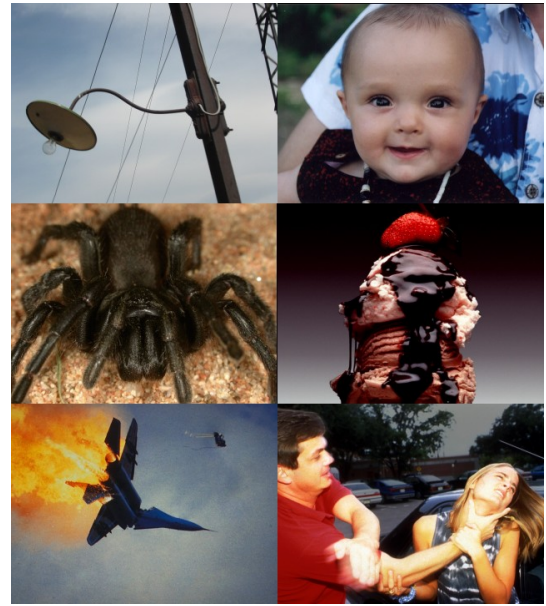

Figure 1. Several representative picture stimuli from GAPED and IAPS databases. Emotion and semantics are specified for each stimulus.

In terms of knowledge management emotionally annotated databases only specify data about the stimuli themselves, but do not describe knowledge stored in the stimuli. Therefore, they should not be confused with databases and taxonomies of knowledge. Also, emotionally annotated databases always contain multimedia documents while knowledge databases do not. On the other hand, knowledge databases describe semantic concepts in detail while emotionally annotated databases only reference semantic labels present in stimuli. The full expression of semantic labels is never found in these databases. It is up to the user to find the most appropriate knowledge database and reasoning engine to deduce the meaning of multimedia semantic descriptors.

All this clearly demonstrates diversification and content variability of currently available emotionally annotated databases. Today these databases are described loosely and informally, they are domain dependant and with arbitrary custom structures, new stimuli cannot be simply added in a database but require a separate set of affective ratings to be acquired through experimentation; existing stimuli must be extracted manually which is frequently tiresome and work-intensive; etc. In light of these drawbacks much can be done to improve functionality and usefulness of existing emotionally annotated databases. One of the most pressing tasks is automatization of stimuli insertion. Exploitation of semantic and affective coupling may greatly contribute in this process.

## III. DEFINING SEMANTIC AND AFFECTIVE COUPLING

Multimedia stimuli have semantics, affect and context. Any semantics will induce an emotion, or affect, in an exposed subject. We are interested in interaction of semantics and emotions since their relationship is not random but subjected to personal knowledge, personal experience, cultural conditioning, collective memory and other similar factors which cumulatively direct articulation of particular semantics towards a specific set of emotions. Difference in articulation of images through pixel-defined content and semantic content is known as semantic gap [17][18]. Here we use the phenomena of semantic gap and broaden its scope on emotional content within the context of affective computing. We define coupling of semantics and affect in emotionally annotated multimedia documents as a deterministic interaction between semantics of a document and affect which its semantics is inducing.

Dimensional model of emotion [13] describes each emotion as a tuple $e_i = \{val, ar, dom\}$ where $val$, $ar$, $dom$ are continuous variables representing valence, arousal and dominance emotional dimensions: $val \in [1,9] \in Val$, $ar \in [1,9] \in Ar$, $dom \in [1,9] \in Dom$. Dominance is frequently omitted from description of emotion space because it was shown to be the least informative measure of the elicited affect [3]. Therefore, for all practical purposes a single multimedia document from an emotionally annotated database can be represented as a coordinate in a two-dimensional emotion space $\Omega_{Emo} = Val \times Ar$. Semantics represents a list of document's objects and events, as well as constraints and functional dependencies between objects and events. These are generally described with free-text keywords (i.e. tags), but more expressive methods may be used such as WordNet synsets or ontology concepts (example [19]). Regardless of representation method we define semantics $s_i$ as a set of descriptors $d_n$ in semantics space $\Omega_{Sem}$

$$s_i = (d_1, d_2, \ldots \quad, s_i \in \Omega_{Sem} \quad (1)$$

Then a surjective functions $f_{se}$ maps elements in semantics space to elements in emotion space $f_{se}: \Omega_{Sem} \rightarrow \Omega_{Emo}$ (Fig. 2).

We introduce semantic distance operator $d_{Sem}$ and emotion distance operator $d_{Emo}$

$$d_{Sem}: \Omega_{Sem} \times \Omega_{Sem} \rightarrow \mathbb{R}$$
$$d_{Emo}: \Omega_{Emo} \times \Omega_{Emo} \rightarrow \mathbb{R} \quad (2)$$

Since $\Omega_{Emo}$ is Euclidian space we can write $d_{Emo} = \sqrt{(val_i - val_j)^2 + (ar_i - ar_j)^2}$ for any two emotions $e_i = (val_i, ar_i)$, $e_j = (val_j, ar_j)$.

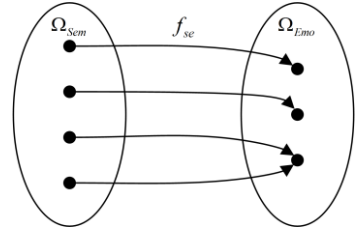

Figure 2. Surjective function $f_{se}$ mapping semantics space domain $\Omega_{Sem}$ to emotion space codomain $\Omega_{Emo}$.

Distance between semantic descriptors can be calculated if they are represented as nodes in a graph or semantic network like WordNet. Indeed, many measures of term similarity exits and can be applied to assess the semantic distance (as example see Semantic Similarity Retrieval Model[20]). Both distance operators have symmetry and can be defined with emotional similarity $sim_{Emo}$ and semantic similarity $sim_{Sem}$.

$$d_{Emo}(e_i, e_j) = d_{Emo}(e_j, e_i) \,, \, e_i, e_j \in \Omega_{Emo}$$
$$d_{Sem}(s_i, s_j) = d_{Sem}(s_j, s_i) \,, \, s_i, s_j \in \Omega_{Sem} \quad (3)$$
$$d_{Emo} = 1/sim_{Emo}$$
$$d_{Sem} = 1/sim_{Sem}$$

The semantic-affective coupling tells us that similar semantics in $\Omega_{Sem}$ induces similar emotion in $\Omega_{Emo}$. Formally the coupling exists between multimedia documents $doc_i = \{e_i, s_i\}$, $doc_j = \{e_j, s_j\}$ with different semantics $s_i \neq s_j$ if $e_i$ in a neighborhood $\varepsilon_{Emo}$ of $e_j$ and $s_i$ is in a neighborhood $\varepsilon_{Sim}$ of $s_j$

$$d_{Emo}(e_i, e_j) \leq \varepsilon_{Emo} \,, \, e_i, e_j \in \Omega_{Emo}$$
$$d_{Sem}(s_i, s_j) \leq \varepsilon_{Sim}, s_i \neq s_j \quad s_i, s_j \in \Omega_{Sem} \quad (4)$$

For example, if we have three distinct images $doc_1 = \{e_1, s_1\}$, $doc_2 = \{e_2, s_2\}$, $doc_3 = \{e_3, s_3\}$, the coupling will exist between $doc_1$ and $doc_2$ only if

$$d_{Sem}(s_1, s_2) \leq \varepsilon_{Sim}$$
$$d_{Sem}(s_1, s_2) < d_{Sem}(s_1, s_3) \leq d_{Sem}(s_2, s_3) \quad (5)$$

and

$$d_{Emo}(e_1, e_2) \leq \varepsilon_{Emo}$$
$$d_{Emo}(e_1, e_2) < d_{Emo}(e_1, e_3) \leq d_{Emo}(e_2, e_3) \quad (6)$$

If (5) is true but (6) is not then $doc_1$ and $doc_2$ are not coupled. This is illustrated in the next figure.

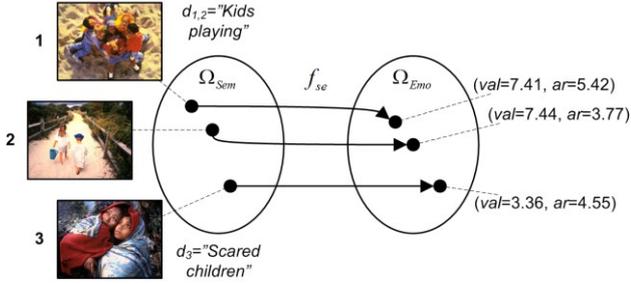

Figure 3. Semantic and affective coupling between images 1 and 2. Image 3 is not coupled with the other two images.

As can be seen in the example multimedia documents $doc_1$ and $doc_2$ are semantically and affectively coupled since their semantic and emotional distance falls within the defined neighborhoods in $\Omega_{Emo}$ and $\Omega_{Sem}$.

## IV. COUPLING EXAMPLES IN EMOTIONALLY ANNOTATED IMAGES

We present two unequivocal examples of semantic and affective coupling in image stimuli from IAPS. Since IAPS semantic descriptors (i.e. keywords) are not interconnected or supervised [15] we had to develop our own software tool Stimulus Generator (StimGen) for research in emotionally annotated databases and construction of well-defined personalized stimuli sequences [21][22][23]. In StimGen we have expanded IAPS and IADS databases to include tags clouds. We use tag clouds in these examples to retrieve stimuli with identical or sufficiently similar semantics as defined in the previous chapter.

The first example analyzes different images of animals in IAPS. We have extracted 40 images with 26 keywords and grouped them into six distinct sets: Dead animals, Dangerous animals, Revolting animals, Neutral animals, Wild animals and Miscellaneous animals. (Fig. 4).

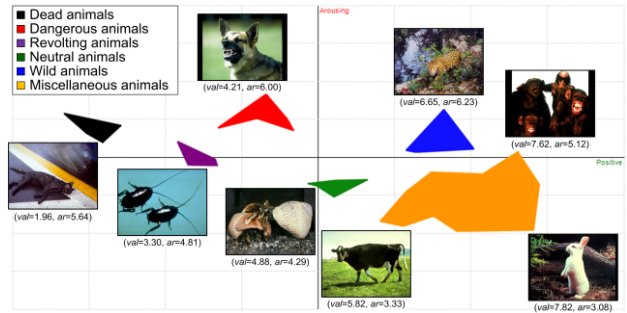

Figure 4. First example of semantic and affective coupling with IAPS images of various animals.

Since IAPS keywords are inadequate in relying true image semantics we had to construct our own specific tag clouds. Sets contain semantically similar images and sets' different semantics do not overlap in the emotion plane. It is immediately apparent that the tag clouds have well-defined boundaries and that they induce distinct sets of emotions. Since each of this emotion sets is elicited with images of similar semantics, we can say that the sets are semantically and affectively coupled.

In the second example (Fig. 5) we select total of 72 stimuli from IAPS tagged with food-related keywords (Beer, Brownie, Cake, Candy, Cheeseburger, Chicken, Desserts, Food …) annotating 35 stimuli, 24 nature-related stimuli (Cave, Desert, Farmland, Mountain, Mountains, Nature, Sunset, Waterfall, …) and 13 stimuli representing various high-action sports (CliffDivers, Motorcyclist, Rafting, SkyDivers, Skysurfer, WaterSkier, Windsurfers, …). These stimuli groups also outline distinctive areas in $Val \times Ar$ plane. From group dispositions it is immediately apparent that their semantics and affect are coupled; groups are small, closed and well-defined.

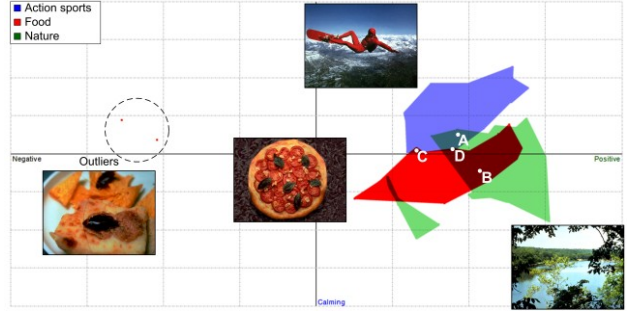

Figure 5. Second example of semantic and affective coupling with IAPS images of food, nature and action sports.

Emotions in points A, B and C in Fig. 5 can be induced with actions sports or nature, food or nature and action sports and food semantics, respectively. The point D emotion (val~6.82, ar~5.12) can be elicited with all three semantics groups (i.e. IAPS tag clods) in this example: food, actions sports and nature. We also note two outlier images with semantics of revolting food. In IAPS these images are tagged with "food" keyword like other non-repugnant images of food. This is another example of how semantically insufficient IAPS tags are and a clear indicator how this emotionally annotated database should be improved.

These examples clearly demonstrate the existence of semantic and affective coupling in emotionally annotated databases' multimedia stimuli. While emotional similarity can be easily established in the dimensional model of emotion which is essentially a simple geometric 2-D/3-D model, this is not also true for the semantics. Our continuing research is focused on using lexical ontologies like WordNet [24], and more formal and expressive upper ontologies such as SUMO [25] to better describe stimuli content and to measure semantic similarity of concepts describing stimuli content.

## V. COUPLING IN SUPPORTED CONSTRUCTION OF EMOTIONALLY ANNOTATED DATABASES

Supported construction of emotionally annotated databases is a much needed feature leading to their more frequent everyday utilization. Since current construction processes require assessment of a large number of individual ratings collected in stimuli annotation experiments, the amount of time, resources and workload required to perform these tasks hider database deployment. Therefore, any type of automatization in stimuli annotation is needed to facilitate the development of emotionally annotated databases and intensify their subsequent use.

The coupling of semantics and affect enables a priori knowledge of stimulus semantics to derive stimulus emotion using semantics and emotion information of stimuli already in the emotionally annotated database. The database construction algorithm may use the coupling to retrieve a set of the most probable emotion annotations for a new stimulus. The results are estimated, sorted by estimation likelihood and displayed to the user. Through interactive user feedback the classification may be optimized in a sequence of question-answer sessions between the stimulus annotation system and human annotation expert. An algorithm for supported and iterative construction of emotionally annotated databases using semantic and affective coupling is given in the figure below.

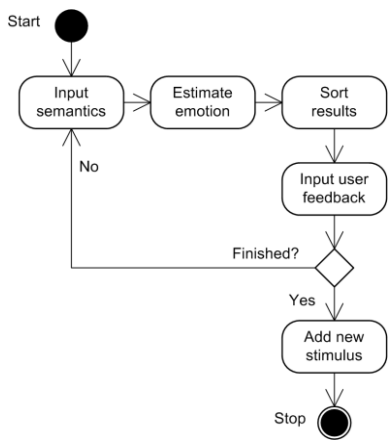

Figure 6. Algorith for supported construction of emotionally annotated databases using semantic and affective coupling.

For each new multimedia stimulus $doc_n = \{e_n, s_n\}$ that is being added to an emotionally annotated database the algorithm's input is $doc_1, doc_2, \ldots \wedge s_n$ and output $e_n = \{val, ar\}$. In other words, the algorithm based on the semantic-affective coupling uses new stimulus' semantics and existing knowledge about semantics and emotion of stimuli already in the emotionally annotated database to assess the most likely classification of new stimulus' emotion.

The proposed algorithm could substantially accelerate creation of new emotionally annotated databases or modification of the existing ones because semantics is more easily defined than emotion. Also a number of content-based or concept-based image recognition procedures could be used to automatically identify, at least partially, the semantics of an image stimulus.

## VI. CONLUSION

We have shown that despite semantic gap semantic-affective coupling is a real phenomenon in emotionally annotated multimedia databases and proposed the algorithm for its utilization. The coupling can fill missing stimuli annotation data using existing information on semantics and emotion. As such it may be envisioned as a method for faster and simpler construction of emotionally annotated databases.

The biggest obstacle in implementation of an efficient coupling estimation algorithm is ambiguous and insufficient description of stimuli semantics in existing emotionally annotated databases. To address this problem our research focuses on applying formal knowledge representation methods for description of image content.

In the continuation of our work we are building core Stimulus Ontology for formal presentation and integration of semantics, affect and context information in multimedia stimuli. We will use WordNet lexical ontology and SUMO upper ontology to reach an optimal description of stimuli content and explore benefits of the coupling.